\documentclass[preprint, raferee]{aastex}
\usepackage{epsfig,amsmath,amsfonts,amssymb,graphicx,subfigure,enumitem}

\usepackage{tabu}
\renewcommand{\arraystretch}{1.0}

\usepackage{bm}
\begin{document}

\title{Evolution of Magnetic Rayleigh-Taylor Instability into the Outer Solar Corona and Low Inter Planetary Space}
\author{Sudheer K.~Mishra}
\affil{Department of Physics, Indian Institute of Technology (BHU), Varanasi-221005, India.}

\author{Talwinder Singh}
\affil{Department of Space Science, University of Alabama in Huntsville, Huntsville AL 35899,USA}

\author{P.~Kayshap}
\affil{UMCS, Group of Astrophysics, UMCS, ul. Radziszewskiego 10, 20-031, Lublin, Poland}

\author{A.K. Srivastava}
\affil{Department of Physics, Indian Institute of Technology (BHU), Varanasi-221005, India}

\begin{abstract}
We analyze the observations from Solar TErrestrial RElations Observatory (STEREO)-A\&B/COR-1 of an eruptive prominence in the intermediate corona on 7 June 2011 at 08:45 UT, which consists of magnetic Rayleigh-Taylor (MRT) unstable plasma segments. Its upper northward segment shows spatio-temporal evolution of MRT instability in form of finger structures upto the outer corona and low inter-planetary space. Using method of Dolei et al.(2014), It is estimated that the density in each bright finger is greater than corresponding dark region lying below of it in the surrounding intermediate corona. The instability is evolved due to wave perturbations that are parallel to the magnetic field at the density interface. We conjecture that the prominence plasma is supported by tension component of the magnetic field against gravity. Using linear stability theory, magnetic field is estimated as 21-40 mG to suppress growth of MRT in the observed finger structures. In the southward plasma segment, a horn-like structure is observed at 11:55 UT in the intermediate corona that also indicates MRT instability. Falling blobs are also observed in both the plasma segments. In the outer corona upto 6-13 solar radii, the mushroom-like plasma structures have been identified in the upper northward MRT unstable plasma segment using STEREO-A/COR-2. These structures most likely grew due to the breaking and twisting of fingers at large spatial scales in weaker magnetic fields. In the lower inter-planetary space upto 20 solar radii, these structures are fragmented into various small-scale localized plasma spikes most likely due to turbulent mixing.
\end{abstract}

\section{Introduction}\label{sec:intro}

Prominences are the most common and well studied large-scale features of the solar atmosphere. They are mainly classified as active region, intermediate, and quiescent prominences. Active region prominences occur around the sunspots over the polarity inversion lines (PILs) in the plage regions (Mackay et al. 2008; Martin 1998). Main features of AR prominences are their long thin threads that occur in the horizontal bundles. Intermediate prominences are evolved outside the active regions, generally in the mid latitude regions between the remnant plages (Mackay et al. 1997). The basic features they consist in form of the multiple thread like structures, which are shorter in length and do not occupy entire PIL (Luna et al. 2012). The quiescent prominences are generally evolved at higher latitudes and associated with the weaker photospheric magnetic fields (Priest 1989). The spatial scale of the quiescent prominence is larger and elongate upto the height of 35-50 Mm into the solar atmosphere. Quiescent prominences have an interesting feature showing the plasma downfall within themselves (Berger et al. 2008; Schmieder et al. 2010). The common phenomenon in these quiescent prominences is the formation of bubbles which have been recently observed (Berger et al. 2010, Hillier et al. 2011). Plasma bubbles, fingers and other small scale structures are not yet observed in active regions and intermediate prominences (Berger et al. 2014). However, these structures provide the information on the magnetic field configuration as well as plasma dynamics associated with various prominences.

The Rayleigh-Taylor instability and other magnetohydrodynamic (MHD) instabilities are well studied phenomena in the laboratory, fusion and astrophysical plasma (e.g., T{\"o}r{\"o}k \& Kliem 2005; Srivastava et al.; 2010, 2013, Foullon et al. 2011, Innes et al. 2012, Moser \& Bellan, and references cited therein). Rayleigh-Taylor instability is first studied by Rayleigh in (1883) and Taylor in (1950). Taylor et al. (1950) have examined small disturbances at the interface of the fluid wall and applied the linear stability theory (Chandrasekhar 1961) to discuss their exponential growth at the fluid interface. The RTI appears when denser plasma is supported against the gravity. 3D numerical simulation have been performed to investigate the evolution of MRT instability. Due to the MRT instability, the formation of localized structures occurs along the magnetic fields (Stone et al. 2007). Filamentary structure of the magnetized plasma under the influence of MRT and  mushroom-like structures are developed in weak magnetic fields (Isobe et al. 2005). Quiescent prominence bubbles were initially observed by Stelmacher et al. (1973), which were later found to be RT unstable (Ryutova et al. 2010). These plasma bubbles were present under the denser prominence material forming MRT instability regions. MHD simulation of prominence based on Kippenhehan and Schlutter model have shown that MRT instability  triggers when a hot denser plasma is introduced into a cooler plasma (Hillier et al. 2011). Prominence bubble analysis shows that the flow of plasma along the bubble's boundary {\bf develops} Kelvin-Helmholtz instability forming its ripples. These act like initial perturbations and Rayleigh-Taylor instability evolves (Berger et al. 2010; Hillier et al. 2011). Simulation for Kippenhahn-Schluter prominence model are recently performed in the frame-work of MHD (Hillier et al. 2011,2012). They described that the formation of upflows arise due to 3D magnetic Rayleigh-Taylor instability. Berger et al. (2010) have observed the upflows in quiescent prominence and consider that these upflows are due to the magnetic Rayleigh-Taylor instability as the interchange or mixed mode of perturbation. Horns, fingers, spikes and plasma bubbles are the small scale structures in the quiescent prominences, which are subjected to the magnetic Rayleigh-Taylor instability (Innes et al. 2012).

One of the most significant event observed by Solar TErrestrial RElations Observatory (STEREO) is a prominence eruption occurred on 7$^{th}$ June 2011. Fast coronal mass ejection (CME) associated with an M2.5 class solar flare and a dome shaped (EUV) wave-front, were erupted out from AR11226 and AR11227 (Cheng et al. 2012). The coronal magnetic reconnection with the field nearby these two active regions triggered the massive prominence eruption and related plasma dynamics (Gesztelyi et al. 2014). Flux cancellation near the two active regions (AR 11226 and AR 11227) have been found as a trigger of this filament eruption as observed by the Heliospheric Magnetic Imager (Yardley et al. 2016). The downfall of the plasma material in filaments possesses arcs, spikes, fingers and horns, which were subjected to the Rayleigh-Taylor instability (Innes et al. 2012). The plasma blobs and fingers are formed due to the magnetic Rayleigh-Taylor instability  during the density evolution of returning plasma blobs in the filamentary structure as observed in the inner corona (Carlyle et al. 2014). Energy release of the impacting prominence has also been observed by different EUV filters of SDO and STEREO (Gilbert et al. 2013). The measurement of drag force acting on plasma fingers during the density evolution of returning plasma have also been estimated into intermediate corona using STEREO/COR-1 data (Dolei et al. 2014). 

As stated above the eruption was started on 7$^{th}$ June 2011 at 06:00 UT and moved towards the inner corona. The full development of this eruptive prominence has been observed  by STEREO-A \& B coronagraphs (COR-1 \& COR-2) and Heliospheric imager (HI-1) in the outer corona and low inter-planetary space respectively. The spatio-temporal scale for this eruption is very large, and it is subjected to the magnetohydrodynamic magnetic Rayleigh-Taylor (MRT) instability. We mark a bulky plasma segment (upper northward part) associated with the observed  prominence eruption in the intermediate corona (Fig.~1). We study this structure during its upward motion and formation of the plasma fingers. We, therefore, investigate the spatio-temporal evolution of the magnetic Rayleigh-Taylor instability in the intermediate corona in plasma segment of this  eruptive prominence. We use the Dolei et al. (2014) method to estimate the  mass density at the location of evolved fingers in the MRT unstable region in the STEREO/COR-1 field-of-view. Using the properties of magnetic Rayleigh-Taylor instability, we measure the magnetic field embedded within the plasma material that is required to suppress the growth of instability in this region. We use STEREO-A/COR-2 and STEREO-A/HI-1 observation to understand the evolution of MRT instability in outer corona (2.5-15 solar radii) and in the low inter-planetary region (15-90 solar radii) respectively. While we complement the observations of MRT instability in the inner corona by Innes et al. (2012) and extend their work, to the best of our knowledge the present work is the first effort to understand the evolution of MRT instability in the outer corona and low inter-planetary space. Moreover, in the lower southward plasma segment of this eruptive prominence, we observe horn-like structure at 11:55 UT in the intermediate corona which also indicates the growth of MRT instability. This type of structure has already been observed in the inner corona in MRT unstable regions by Innes et al. (2012). Falling plasma blobs are also evident in both the plasma segments. In Sect. 2, we discuss the observations and data analysis. Observational results are outlined in Sect. 3. Discussion and conclusions are presented in the last section.

\begin{figure}
\caption{ Full disk image of COR-1 onboard STEREO-A \& B have captured the plasma segment of an eruptive prominence at 7 June 2011. Fingers are clearly visible at 08:45 UT on this upper northward MRT unstable segment in COR-1 FOV. }
\mbox{
\hspace{-3.0cm}
\includegraphics[scale=0.8,angle=90,width=14.0cm,height=14.0cm,keepaspectratio]{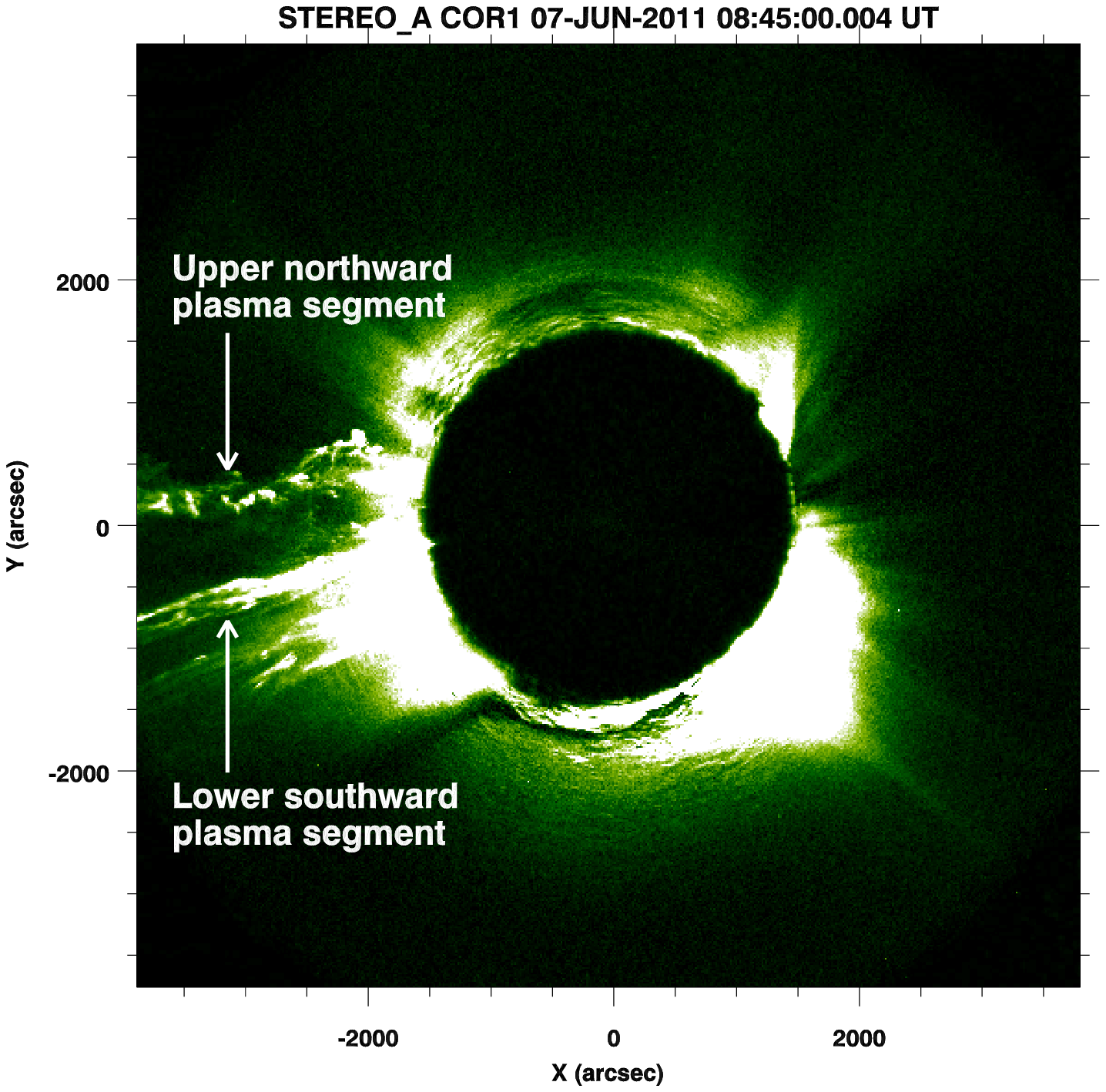}
\hspace{-5.4cm}
\includegraphics[scale=0.8,angle=90,width=14.0cm,height=14.0cm,keepaspectratio]{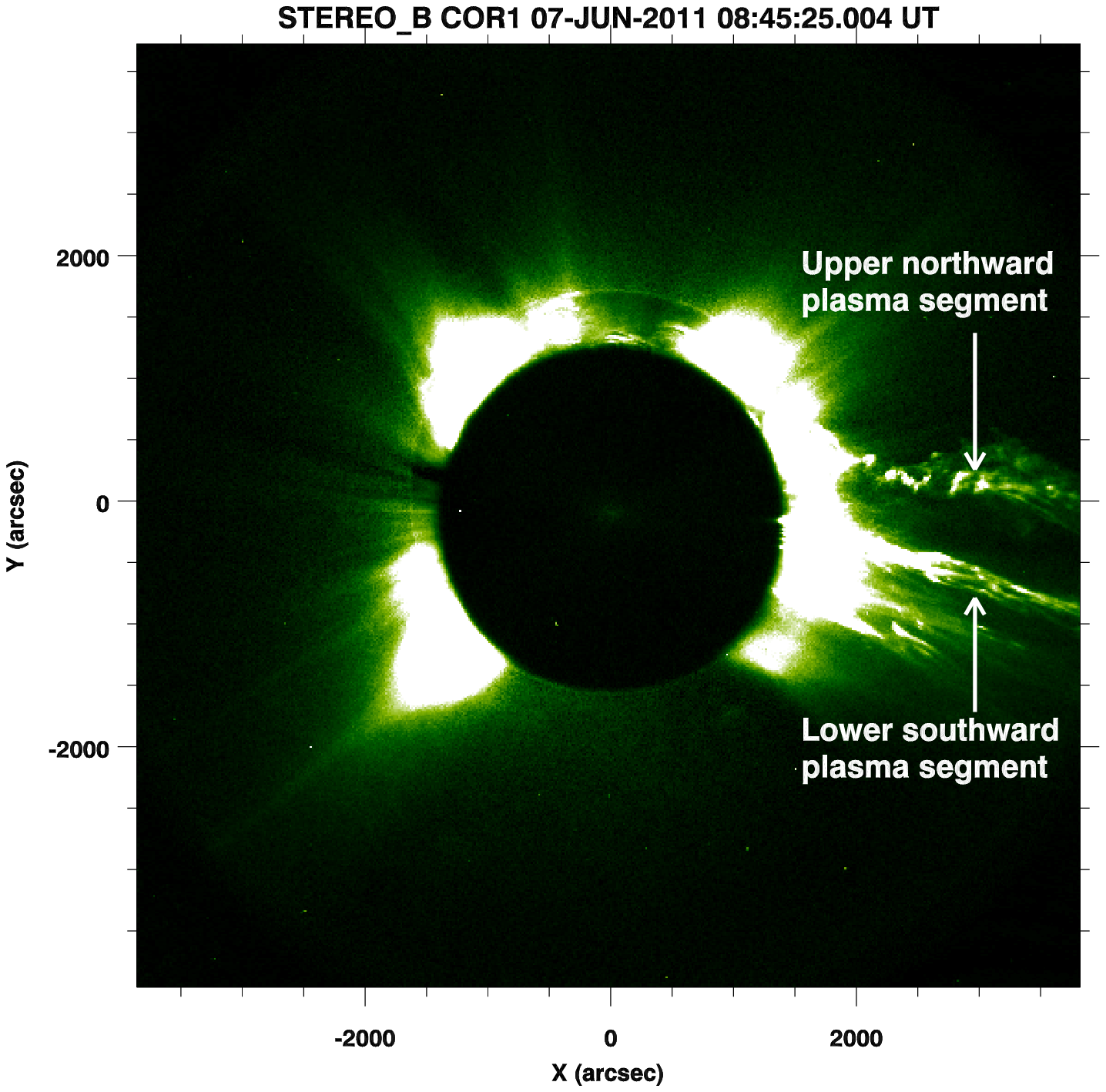}
}
\end{figure}

\section{Observational Data and Analysis} 
In the present paper, we study the observational data obtained by two white-light coronagraphs (COR-1 \& COR-2, Howard et al. 2008) and Heliospheric Imager (HI-1, Howard et al. 2008) onboard STEREO-A spacecraft. The gradient of coronal brightness is very large in corona, therefore, we use COR-1 to observe the intermediate corona, while outer corona (2.5-15 solar radii) is observed by COR-2. COR-1 has a field of view of 1.5 to 4.0 solar radius. The data taken by COR-1 has minimum cadence of 5 minutes, and after each five minutes three consecutive images have been captured with time interval of 9 seconds and at every 120 degree angle ($0\,^{\circ}$, $120\,^{\circ}$, $240\,^{\circ}$) to observe the polarized brightness (pb) sequence.

A prominence is erupted from NOAA AR 11226/11227 on 7$^{th}$ June 2011 starting at 06:00 UT. It was subjected to the magnetic Rayleigh-Taylor (MRT) instability in the inner corona (Innes et al. 2012). Later it splits in various plasma segments in the intermediate corona, while the upper part is already ejected out. In the present work, we observe the evolution of MRT instability in the plasma segment lying in the upper northward part (Fig. 1) of this unbounded eruptive prominence system in the intermediate and outer part of the corona and low inter-planetary space. The temporal evolution of the MRT unstable finger structures is shown in Fig. 2. Some falling plasma blobs are also evident in the lower part of this plasma segment at later times. We study the spatially evolved structures in the MRT unstable region in coronagraphic (COR-1 \& 2) field-of-views. We observe that these finger structure are fully evolved at 08:45 UT on 7 June 2011 in plasma segment as seen in COR-1 (Fig.~3). In the lower southward plasma segment of this eruptive prominence in the intermediate corona, horn-like structure is  clearly visible at 11:55 UT which is associated with the falling plasma blobs (Fig.~2).
We use COR-1 full disc image having 512 x 512 pixels$^{2}$ FOV with the spatial resolution of 7.5"/pixel.

It should be noted that the identified finger structures are also visible in STEREO-B/COR-1 FOV from a different angle. STEREO-B COR-1 observations also show the formation of the localized plasma chunks as finger-like perturbations due to MRT instability (Fig.~2). Moreover their dynamical evolution does not yield any large scale helical motion, instead it shows the formation of localized fingers as displayed in the STEREO-A\&B/COR-1 FOV (Fig.~2). Formation process of these localized finger structures as MRT instability is discussed in detail in subsection 3.1.

COR-2 has a field of view of 2.5 to 15 solar radius (Howard et al. 2008). The COR-2 has a minimum cadence of 15 minutes. These MRT unstable structures have been seen into the outer corona also in COR-2 FOV (Fig.~4). The MRT unstable structures (shown by dotted red arcs in Fig.~4) are fully evolved at 14:39 UT for the same eruptive plasma segment when it completely spread in the COR-2 FOV. Mushroom-like/omega-shaped MRT unstable structures are visible into the outer corona between 6-13 solar radii. These are large scale structures in the outer corona which are best seen in STEREO-A/COR-2 observations..

The lower inter-planetary region is observed by STEREO-A/HI-1 (Howard et al. 2008). HI-1 observes visible light from the outer part of COR-2's field-of-view to almost one third of the distance to Earth's orbit. i.e. from 15 to 80 solar radii. The HI-1 has a minimum cadence of 40 minutes. We have taken an image on 8 June 2011 at 00:49 UT when the unstable plasma segment is completely evolved in the HI-1 FOV. The plasma segment, which was MRT unstable upto 15 solar radii, becomes fragmented in form of localized plasma spikes in half of the FOV as observed by HI-1.

The basic calibration and background removal of COR-1, COR-2 and HI-1 data have been performed using Solarsoft IDL programs "secchi\_prep.pro"\footnote[1]{https://hesperia.gsfc.nasa.gov/ssw/stereo/secchi/doc/secchi\_prep.html} and its associated sub-routines.

\begin{figure*}
\caption{Sub-region of the STEREO-A \& B COR-1 FOV on 7$^{th}$ June 2011 when a MRT unstable plasma segment is in eruptive state. The temporal evolution of MRT instability in form of finger structures and horn (H) have been observed. The animations show the complete evolution of magnetic Rayleigh-Taylor unstable finger {\bf structures}, falling blobs and associated horn-like structure. Some snapshots of these animations have been shown in Fig.~2.}
\mbox{
\hspace{-0.9cm}
\includegraphics[scale=0.8,angle=0,width=18.2cm,height=18.2cm,keepaspectratio]{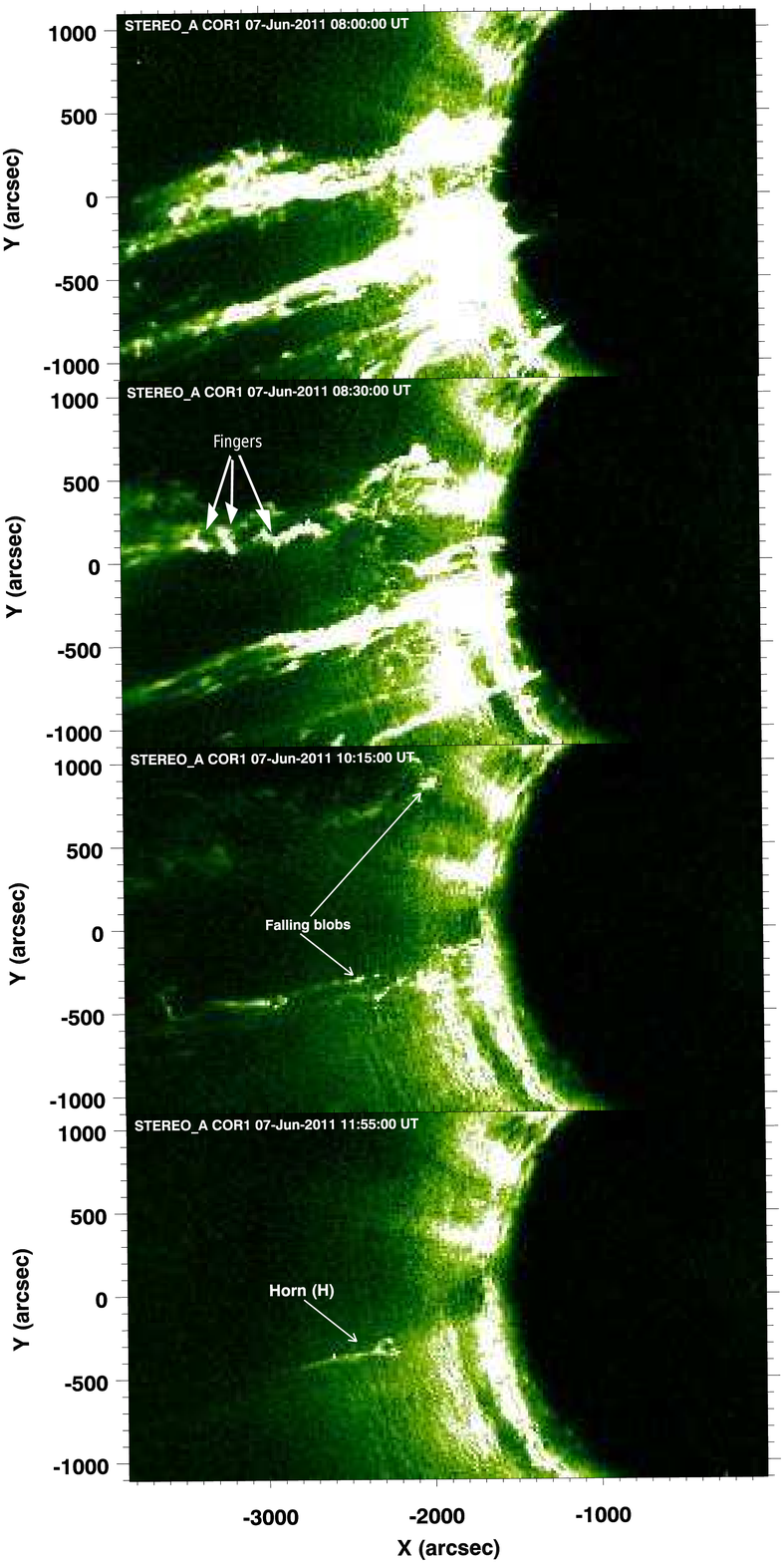}
\includegraphics[scale=0.8,angle=0,width=18.2cm,height=18.2cm,keepaspectratio]{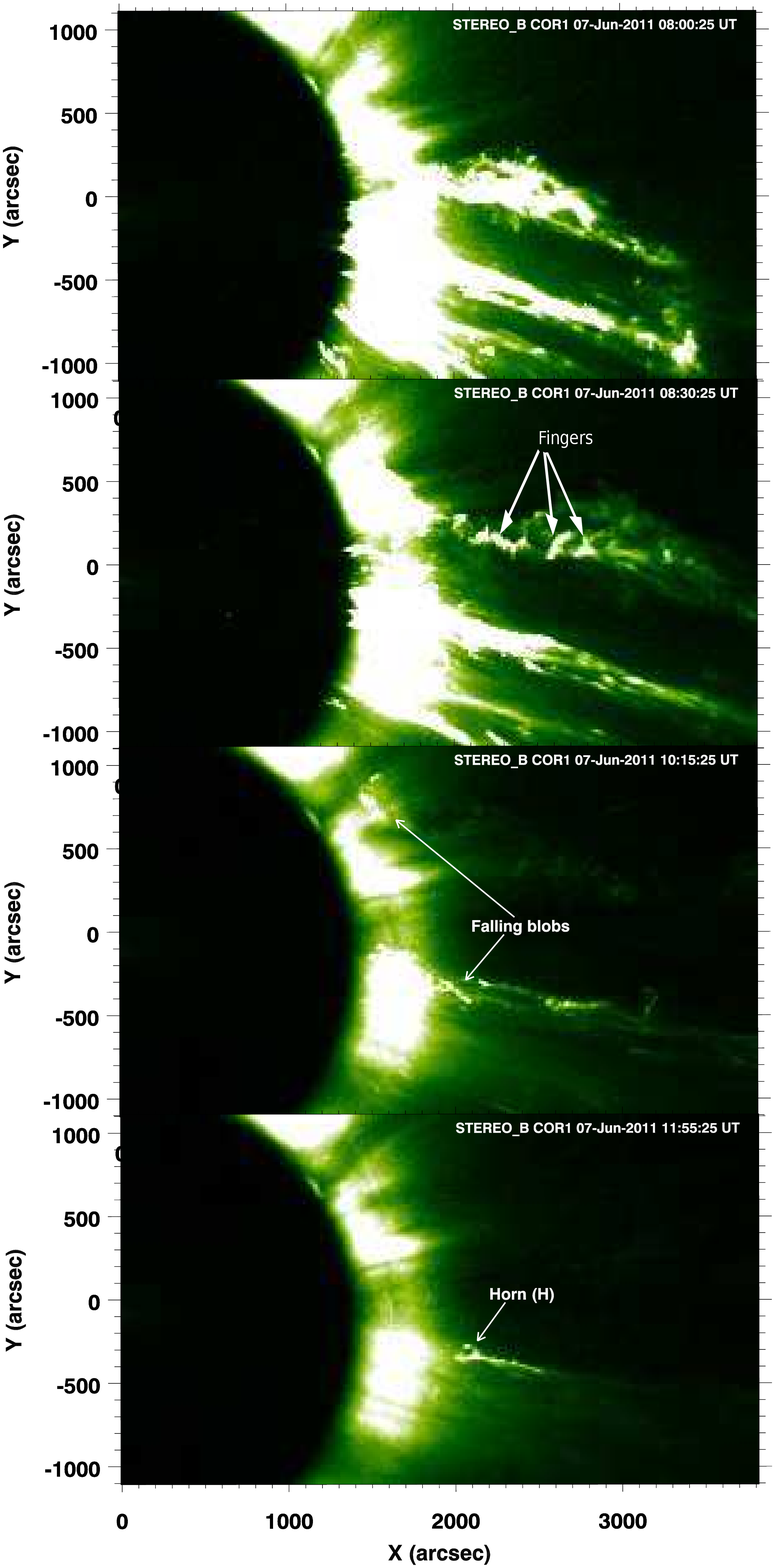}
}
\end{figure*}


\section{Evolution of MRT unstable prominence segments in the intermediate corona}

The prominence eruption has been seen in the intermediate corona on 7 June 2011 during 06:00 UT to 12:00 UT by STEREO-A \& B/COR-1 (Fig.~2). Before appearing into intermediate and outer corona, this prominence in the lower corona (1.05-1.3 solar radii) is observed using high resolution and high cadence temporal image data captured by Atmospheric Imaging Assembly (AIA) onboard the Solar Dynamic Observatory (SDO; Lemen et al. 2012) as reported by Innes et al. (2012). In the lower corona, the magnetic Rayleigh-Taylor unstable structures (e.g. fingers, spikes, horns) are developed in the falling plasma materials as observed by Innes et al. (2012). These ejected plasma move outward due to large scale expansion and create the small scale cavities. These cavities are associated with the prominence eruption site, therefore, a pressure gradient has been developed. Due to expansion of the cavity in the lower corona, localized finger structures are formed within it. These finger structures have been clearly seen in the lower corona (Fig.4; Innes et al. 2012) due to high quality observations from SDO/AIA. 

The slung plasma material associated with this prominence eruption has been moved from inner corona to the intermediate corona (1.4-4.0 solar radii). The intermediate corona is observed by STEREO/COR-1, which has low spatio-temporal resolution as compared to SDO/AIA. The formation of cavities in the intermediate corona has not been clearly seen due to the limitation of the instruments in terms of spatial resolution. However, the bright fingers (F1-F5) and dark plasma regions (B1-B5) have been clearly observed in the northward upper plasma segment of this eruptive prominence (Fig.~3). During the expansion of this part of the prominence, the cavity breaks and cool plasma lies at the top while hot plasma at high-pressure lies underneath the cool erupted prominence. We have selected a sub-region containing eruptive plasma segments in order to observe their temporal evolution in COR-1 FOV (Fig.~2). We have used time-sequence of the images to observe the development of magnetic Rayleigh-Taylor instability in STEREO-A \& B/COR-1 FOV. We have observed that initially the plasma segments are in the form of bulky continuous material. After 08:00 UT, the bottom part of cool plasma starts to fall towards the Sun's surface and its top segment is continuously expanding. After 08:30 UT, the upper northward part of the plasma segment is fragmented into the localized plasma chunks due to the evolution of MRT instability. These MRT unstable finger structures have been fully developed at 08:45 UT into the intermediate corona in STEREO-A \& B FOV. For this observed plasma segment, we identify the fingers (F1-F5; Fig.~3) and dark plasma regions (B1-B5; Fig.~3) associated with MRT unstable plasma (Figs.~2-3). In the  later time during 09:35 UT to 10:30 UT, the full development of falling blobs in upper northward plasma segment can be clearly observed (cf., Fig.~2; animation.gif). At 10:15 UT, all the cool plasma in form of falling blobs (indicated by arrows in animation.gif; Fig.~2) is visible in both the eruptive plasma segments. Upto this time, the plasma containing all the fingers in the upper northward segment of this eruptive prominence has already been moved into the outer corona. In the lower southward plasma segment, the falling blobs consist of another magnetic Rayleigh-Taylor unstable structure namely horn (H). This MRT unstable structure has been clearly observed at 11:55 UT in STEREO-A \& B/COR-1 FOV (Fig.~2). They have already been observed in the lower corona (Fig.~6; Innes et al. 2012) using higher quality data from SDO/AIA. The kinematics of these falling plasma blobs have also been observed in the intermediate corona in the same eruptive prominence segment using STEREO-A \& B/COR-1 data ({\bf Dolei} et al. 2014; Fig. 3).
\begin{figure*}
\caption{A sub-region of the top: STEREO-A \& bottom: STEREO-B COR-1 FOV at a time 08:45 UT on 7$^{th}$ June 2011 when MRT unstable plasma segment is evident. This segment lies in the upper {\bf northward} part of an eruptive prominence. Denser and bright finger structures (F1-F5) are seen in this plasma segment. Darker plasma regions (B1-B5) are also visible just below of these finger structures.}
\vspace{-1cm}
\mbox{\includegraphics[scale=0.8,angle=0,width=20.0cm,height=15.0cm,keepaspectratio,trim=10mm 30mm 10mm 30mm, clip]{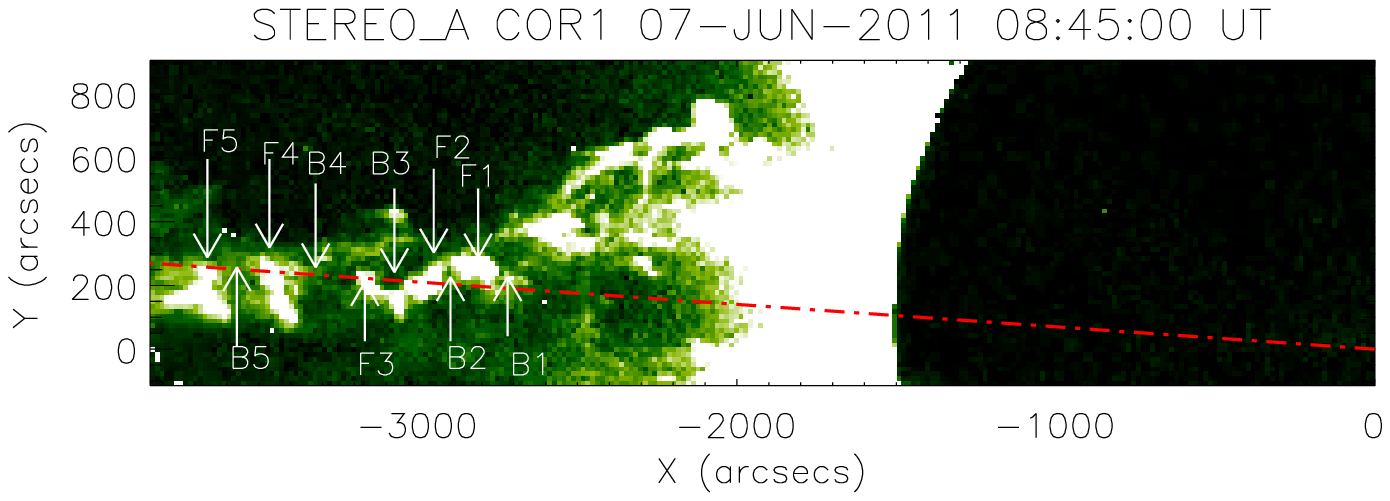}}\\
\vspace{-5cm}
\mbox{\includegraphics[scale=0.8,angle=0,width=20.0cm,height=15.0cm,keepaspectratio,trim=10mm 30mm 10mm 30mm, clip]{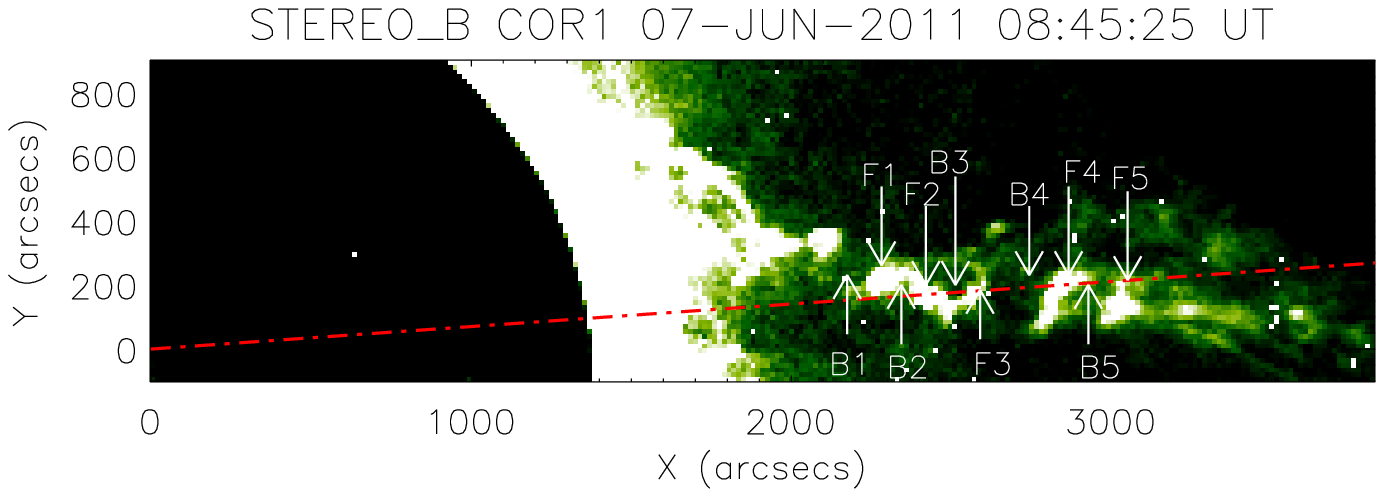}}\\
\end{figure*}

Hillier et al. (2012) studied the nonlinear dynamics of magnetic Rayleigh-Taylor instability in the Kippenhahn-Schluter prominence model. During the upflow, a magnetic Rayleigh-Taylor instability grows and creates rising bubbles and falling spikes. Here, we observe similar dynamical phenomenon during the upflow of prominence with the formation of dense bright finger structures (F1-F5) and less dense dark plasma regions (B1-B5) as seen in STEREO-A \& B/COR-1 FOV (Fig.~3). We assume that the formation of finger structures and dark plasma regions is governed by same physical mechanism as described by Innes et al. (2012) in the lower corona. Magnetic field also plays a crucial role in the growth of MRT instability and finger formation. As the plasma segment expands due to a pressure gradient, the component of Lorentz force reduces the mixing of hot and cool plasma at the perturbation interface, therefore, the outer part of the plasma segment breaks into larger bright finger structures and dark plasma regions  (Fig.~3). In the present case, after the ejection of the upper part of the flux-rope, the northward segment of the unbound prominence system get fragmented and move upward with comparatively lower speed ($\approx$300 km s$^{-1}$) in the intermediate/outer corona (upto 4 solar radii). Moreover, spatio-temporal evolution of these fingers occur in form of localized structures in this segment rather than its fast upward motion and complete ejection like plasmoids.
The MRT instability evolves when two non-viscous fluid of constant density, are separated by a contact of discontinuity perpendicular to gravity with an uniform magnetic field component present at the interface. If the higher density fluid is supported above a lighter fluid, therefore, the interface between two fluid layers became unstable with a perturbation represented by a wave vector k. The gravitational potential energy is converted into kinetic energy and vice-versa, creating highly dense plasma spikes and low denser dark plasma regions below them. In the observed eruptive prominence (Figs.~1-3), we have detected two types of magnetic Rayleigh-Taylor unstable structures, i.e., fingers (in the upper northward plasma segment of the prominence at 08:45 UT; Fig.~3) and the horn (H) (in the lower southward plasma segment of prominence during downfall at 11:55 UT; Fig.~2). We aim to estimate the magnetic field required to suppress the growth of magnetic Rayleigh-Taylor instability in the northward plasma segment using linear stability theory of MRT instability (Chandrasekhar 1961). We have focused on this plasma segment of the prominence, which consists of MRT unstable finger structures.

We assume that the finger structures (F1-F5) are lying in the perpendicular plane w.r.t. the direction towards the Sun's center (red-line in Fig.~3), which is a radial direction. We also assume that the radial-component of magnetic field (along the red-line: Fig.~3) is directed towards the Sun center, and it has no actual effect on suppressing the growth of magnetic Rayleigh-Taylor instability. The radial field works to only hold the vertically extended prominence plasma segment. This is the planar magnetic field (perpendicular to radial direction) which is acting at the interface of the discontinuous surface to suppress the resulting MRT instability. The 3D finger structures (F1-F5: Fig.~3) consist of higher density $\rho_{h}$, whereas the dark regions (B1-B5: Fig.~3) possess the lower density $\rho_{l}$. It is acceptable to assume that the bright denser finger structures (F1-F5) are associated with the cool prominence plasma material and the dark plasma regions (B1-B5) are analogous to the hot coronal plasma materials. The hot and cool plasma regions may be lying on the different planar magnetic field and shouldn't be magnetically connected, {\bf because} the conduction in the corona is along the field lines. Therefore, the planar magnetic field must be parallel to the discontinuous boundary interface between hot and cool plasma layers. The magnetic Rayleigh-Taylor instability occurs if the parallel component of the wave vector is lower than a critical value of a wave vector $k_{c}$, i.e., $|k|< |k_{c}|$.

 In the intermediate corona, in this observed finger structures, the density will be $\rho_{h}$. The density gradient will work inward towards the sun center w.r.t. the below lying dark regions. Therefore, the density (thus pressure) gradient will work inward in the direction of the gravity. The magnetic tension force component of Lorentz force acting outward will hold the localized plasma fingers under stability criteria against the gravity. Similarly all other finger structures will also have the magnetic tension component of Lorentz force against the gravity in the MRT unstable plasma segment (Fig.~3). The tension force associated with the magnetic field suppresses the mixing of lighter and denser material at small scales thus plasma bubbles and fingers are evolved with higher growth rate (Stone et al. 2007). 
\vspace{1cm}
\subsection{Density estimation in MRT unstable fingers}
{\bf Dolei et al.} (2014) describes a method to calculate the electron density in moving solar coronal structures using STEREO/COR-1 brightness images. They derive that electron density ($n_e$, which can be estimated by solving the quadratic equation:
\begin{equation}
1.57\times 10^{-9} W^{-1} f(v) n_e^2+n_e-\frac{tB-tB_{cor}}{<B>_d d}=0 ,
\end{equation}
where, $W=\frac{1}{2}\Big[1-\sqrt{1-\frac{1}{(1+h)^2}}\Big]$ in which 'h' is the height of finger in Solar radii.\\
$tB$ is the brightness found at a pixel location where we are trying to estimate the density in COR-1 total brightness image. $tB_{cor}$ is the brightness contribution by the Thompson scattering of coronal electrons. Dolei et. al (2014) have found that $tB_{cor}$ is 100-1000 times less than $tB$, therefore, we ignore it in the present study. $f(v)$ is a correction factor that needs to be taking into account the Doppler brightness effect in H${\alpha}$ emission. Rompolt (1980) have found that for speeds exceeding 100 km $s^{-1}$ and heights exceeding 1 solar radius, this factor is approximately 4. $<B>_d$ is the average brightness per electron due to Thompson scattering at a given distance from the Sun. We find this value using 'eltheory.pro' in the solarsoft. We need the angle of the finger from the plane-of-sky and it is calculated by using tie pointing method as described by Inhester (2006). This method is a triangulation technique using 2 viewpoints of STEREO-A \& B and is available in Solarsoft under 'scc\_measure.pro'. This angle is also used while calculating the real 3-D distances between fingers in order to estimate the magnetic field.
Finally, $d$ is the depth of the finger along the line of sight. Since this value can not be observed, we approximate it as the size of the finger in the plane of sky perpendicular to the radial line. To convert the number density of electron into the mass density, we assume that the region of interest consists of 90\% of Hydrogen and 10\% Helium. This then results in a mass of $1.97 \times 10^{-24}$ gm/electron. The estimated mass density within the finger structures and below lying dark region are shown in Table.1.

\begin{table}[h!]
\caption{Mass density estimation within the fingers (F1-F5) and in the dark regions (B1-B5) using the method of Dolei et al. (2014) w.r.t. normalized height. The normalized height is considered w.r.t. the position B1 situated at 2073 Mm from Sun's center. The magnetic field is estimated using linear MRT theory which is require to to suppress the growth of instability in the finger structures.}
\label{table:1}
\begin{center}
\begin{tabular}  {|p{2.8cm}|p{2.8cm}|p{2.8cm}|p{2.8cm}|p{2.8cm}|p{2.8cm}|}
\hline
Finger No.& Blob No.& Projected height (Mm) with respect to position 'B1'&  Mass density within finger ($\rho_{h}$)$\times\,10^{-17}$ g cm$^{-3}$& Mass density of below lying dark region ($\rho_{l}$)$\times\,10^{-17}$ g cm$^{-3}$& Magnetic field (mG)\\
\hline
  & B1 & 0 &  & 2.25$\pm$0.23 &  \\ 
\hline
F1 &  & 83.4$\pm$7.4 & 5.96$\pm$0.6 &  & 40.5$\pm$5.39 \\
\hline
& B2 & 104.3 &  & 2.03$\pm$0.2 &  \\ 
 \hline
F2 &  & 194.6$\pm$8.7 & 4.93$\pm$0.49 &  & 34.0$\pm$5.43\\
\hline
 & B3 & 271.1 &  & 1.31$\pm$0.13 &  \\ 
 \hline
F3 &  & 312.8$\pm$6.6 & 4.73$\pm$0.47 &  & 32.8$\pm$3.77\\
\hline
 & B4 & 410.0 &  & 0.76$\pm$0.08 &  \\ 
 \hline
F4 &  & 529.2$\pm$6.4 & 2.17$\pm$0.22 &  & 28.2$\pm$3.49\\
\hline
 & B5 & 611.6 &  & 1.17$\pm$0.12 &  \\ 
 \hline
F5 &  & 695.0$\pm$8.0 & 2.38$\pm$0.24 &  & 21.8$\pm$3.83\\
\hline
\end{tabular}
\end{center}
\end{table}


\subsection{ Estimation of magnetic field in MRT unstable region to suppress the instability}
For a plasma system (here observed fingers in the intermediate corona), when a contact of discontinuity formed where a heavy fluid is supported above a lighter fluid against the gravity, then this boundary is unstable to the perturbations that grow by converting gravitational potential energy into the kinetic energy and creating rising and falling finger structures (Hillier et al. 2016; Fig.~3). If $\gamma$ is the growth rate of Rayleigh-Taylor instability, $\rho_{h}$ and $\rho_{l}$ indicate the upper density and lower density respectively at the interface of finger and below lying darker region. If g is the acceleration due to gravity, the linear stability theory (Chandrasekhar 1961; Priest 2014) describes that the growth rate of Rayleigh-Taylor instability is,
\begin{equation}
\gamma^{2}=-gk{\frac{(\rho_{h}-\rho_{l})}{(\rho_{h}+\rho_{l})}}+{\frac{B^{2}k^{2}cos^{2}\alpha}{2\pi(\rho_{h}+\rho_{l})}} ,
\end{equation} 
where $\alpha$ is the angle between the magnetic field (B) and the wave vector $k$ of the MRT perturbations. If the interface of ideal two inviscid, perfectly conducting fluids separated by a constant discontinuity with an uniform magnetic field B is parallel to the perturbations interface (k), therefore, $\alpha=0$. This particular condition makes second term in Eq.(2) positive, which shows a stabilizing effect against gravity to suppress the growth rate of MRT instability (Jun et al. 1995, Priest 2014, Hillier et al. 2016).
Instability occurs if the wave vector is lower than a critical value of a wave vector $k_{c}$, i.e., $|k|< |k_{c}|$. The magnetic field embedded with the plasma for a given critical wavelength ($\lambda_{c}$) associated with the wave vector ($k_{c}$) to suppress the growth rate (growth rate $\gamma=0$) of magnetic Rayleigh-Taylor instability is (Chandrasekhar 1961; Jun et al. 1995; Ryutova et al. 2010; Priest 2014),
\begin{equation}
B=\sqrt{{g\lambda(\rho_{h}-\rho_{l})}},
\end{equation}
 The acceleration due to gravity depends upon the height as,
\begin{equation}
g_{h}=g_{0}\frac{R_{s}^{2}}{(R_{s}+h)^{2}} ,
\end{equation}
where $g_{0}$ is the surface gravity of the Sun (274 m$s^{-2}$), '$g_{h}$' is the gravity at height 'h' from the solar surface and '$R_{s}$' is the radius of the Sun. In Fig.~3, we estimate the projected height of the fingers in COR-1 FOV. To estimate the real height of the finger structures, we have used tie pointing method as given by Inhester (2006). Using the routing $scc\_measure$.pro available in Solarsoft, we estimate the real height (solar radii) of fingers and plasma blobs. We measure the gravity at the heights in each finger structures. Using Eq. 3 with the estimated gravitational acceleration, density difference $(\rho_{h}-\rho_{l})$ between fingers and below lying dark regions, the characteristics wavelength of the perturbations (i.e., actual distance between two fingers), we estimate the average magnetic field in the MRT unstable plasma segment (see; Table 1). As mentioned above, we use linear analysis of the MRT instability as observed in COR-1 FOV. In the linear regime of the MRT instability, the width of the finger structure must be lying between 0.1$\lambda$ to 0.5 $\lambda$ (Sharp et al. 1984). The width of the finger and the characteristic wavelength of the perturbation could match only in case of non-linear phase of MRT instability, which only arise in the outer corona when density falls-off rapidly (Fig.~4). In the {\bf intermediate} corona, the width of the finger structure is only a fraction of distance between two consecutive fingers which we estimate as a characteristic wavelength ($\lambda$). We also measure that the longitude angle for this plasma segment is $40^{0}$ in COR-1 FOV. We scale the projected wavelength by $\frac{1}{cos(40^{0})}$ to obtain the real wavelength ($\lambda$) in order to estimate the magnetic field (Eq.~3). For a given instant and full field of view, we observed finger structures as indicated in Fig.~3. We have used the angle between wave vector and magnetic field $\alpha=0^{0}$, i.e. maximum value of {\bf cos$\alpha$=1}. It reduces the estimated magnetic field, which can be considered as a lower bound estimated field that may require to suppress the instability. There are always some uncertainty in characteristics wavelength measurements due to the finite length of the fingers and the dynamics of the observed plasma segment in the projected 2-D space. The STEREO-A\&B/COR-1 data have been used to measure the mass density withing the fingers and dark plasma regions using {\bf Dolei} et al. (2014) method. There are always uncertainty to locate the real 3-D location of fingers and dark plasma regions, as well as the angle of the fingers from the plane-of-sky, which are required to measure the density and underlying uncertainty ({\bf Dolei} et el. 2014). 
{\bf Including all these uncertainties, we have estimated the field error below $<18 \%$, which is fairly accurate diagnostics of the magnetic field.}

 We estimate the magnetic field along MRT unstable plasma segment associated with a prominence. This unbounded plasma segment consists MRT finger structures, evolved fully at the time at 08:45 UT in the COR-1 FOV. We notice in Table 1 that the estimated magnetic field along MRT unstable plasma segment varies spatially as (21.8-40.5)$\pm$5.4 mG. This segment is the part of an unbound prominence eruption, whose parts have fallen back within the inner corona (Carlyle et al. 2014), while rest of the part moves through the intermediate corona (Fig.~2) still carrying the evolved MRT instabilities. The diagnosed magnetic field inside these plasma segment are slightly higher than the typical outer coronal magnetic field, which is obvious as these eruptive segment carry their own coronal currents and belong to the part of a prominence (Rust 1967; Priest 2014). It should be noted that we do not aim to estimate the radial profile of the magnetic field in intermediate corona. This is just the diagnostics of magnetic field required to suppress the magnetic Rayleigh-Taylor instability (Ryutova et al. 2010) in the plasma segment in intermediate corona which is MRT unstable. The magnetic field has been also estimated in the same region but in different eruptive plasma segment using STEREO-A/COR-1 data by the consideration  of drag force that lies between (40-60)$\pm$10 mG in the range of 2.0 to 3.7 solar radii (Dolei et al. 2014). The estimated magnetic field values using the observation of MRT instability are of the same order to the typical intermediate-coronal magnetic field values as diagnosed by Dolei et al. (2014). Therefore, the present case confirms the observations of MRT instability in an eruptive plasma segment, and its diagnostic capability in estimating the magnetic field conditions locally.

\section{Behavior of the  plasma segment in the outer corona and in the low inter-planetary space}
The upper northward MRT unstable plasma segment moves into the outer corona and observed by COR-2 on 7$^{th}$ June 2011 at 14:39 UT, around 6 hour later than as observed by COR-1. We have used the temporal sequence of images of STEREO-A/COR-2 to observe the evolution of mushroom-like structures in MRT unstable region (Fig.~4). These structures are indicated by (M1-M5) at a time 14:39 UT on the 7$^{th}$ June 2011 in COR-2 FOV. The region of interest has been shown by dotted box in Fig.~4 (right-panel) and its sub-map is plotted into the left side that clearly display the MRT unstable mushroom-like structures (as indicated by red arcs).
\begin{figure*} 
\caption{ Right: The temporal sequence of images which shows the erupting upper northward plasma segment in the STEREO-A/COR-2 on 7 June 2011. The evolution of the mushroom-like structures in the outer corona are observed in the plasma segment. A box has drawn to indicate the region of interest. Left: The box region is displayed to identify the mushroom-like MRT unstable structures.}
\vspace{1.0cm}
\includegraphics[scale=0.8,angle=0,width=18.0cm,height=18.0cm,keepaspectratio]{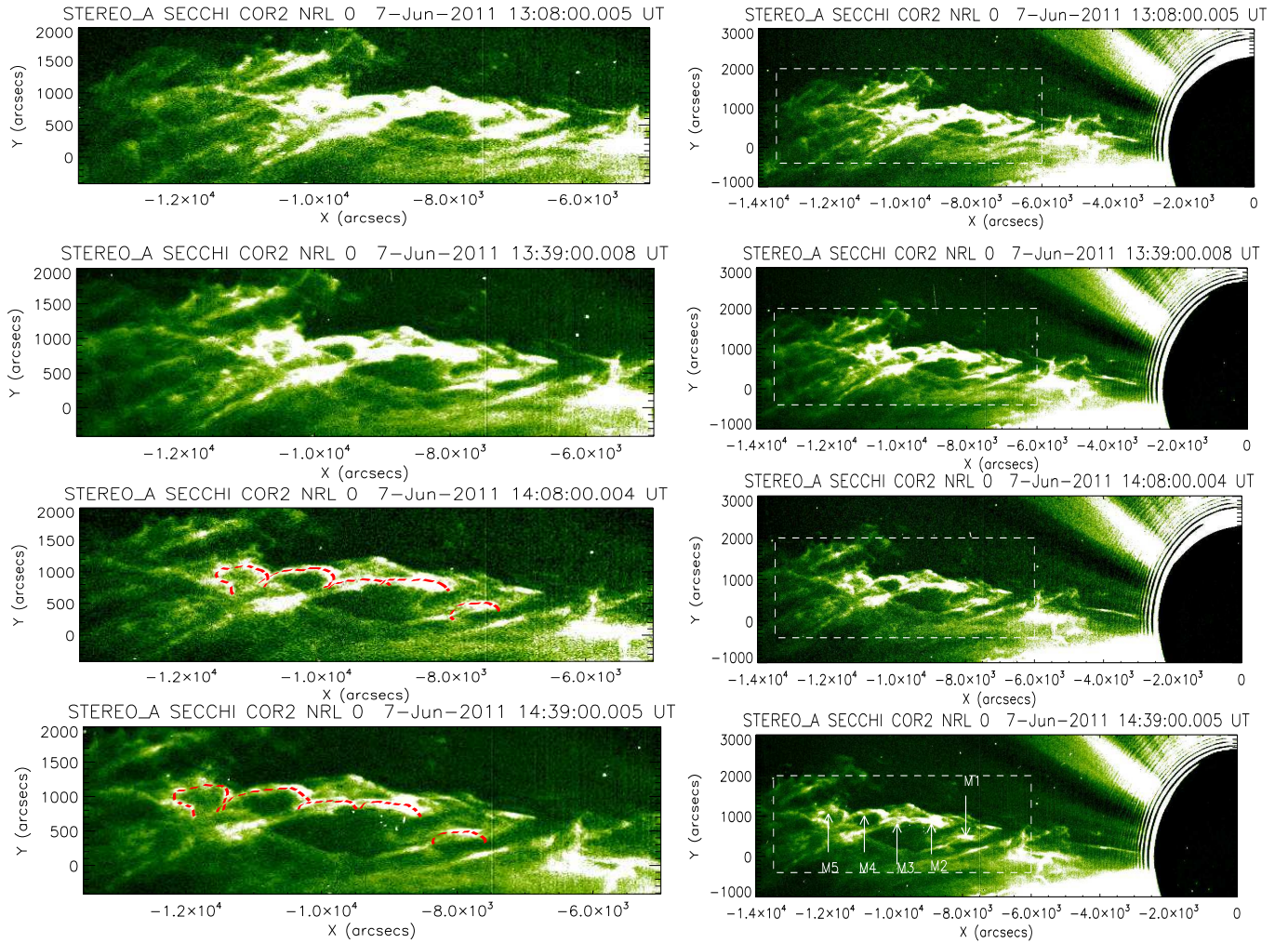}
\end{figure*}
\begin{figure*}
\caption{Right: An STEREO-A/HI-1 image to observe the lower inter-planetary region on 8 June 2011 at 00:49 UT. The eruptive plasma segment is fully extended upto 20 solar radii. Left: Small scale fragmented plasma structures are clearly visible in the inter-planetary space.}
\mbox{
\includegraphics[scale=0.8,angle=0,width=9.5cm,height=10.0cm,keepaspectratio,trim=5mm 5mm 5mm 5mm,clip]{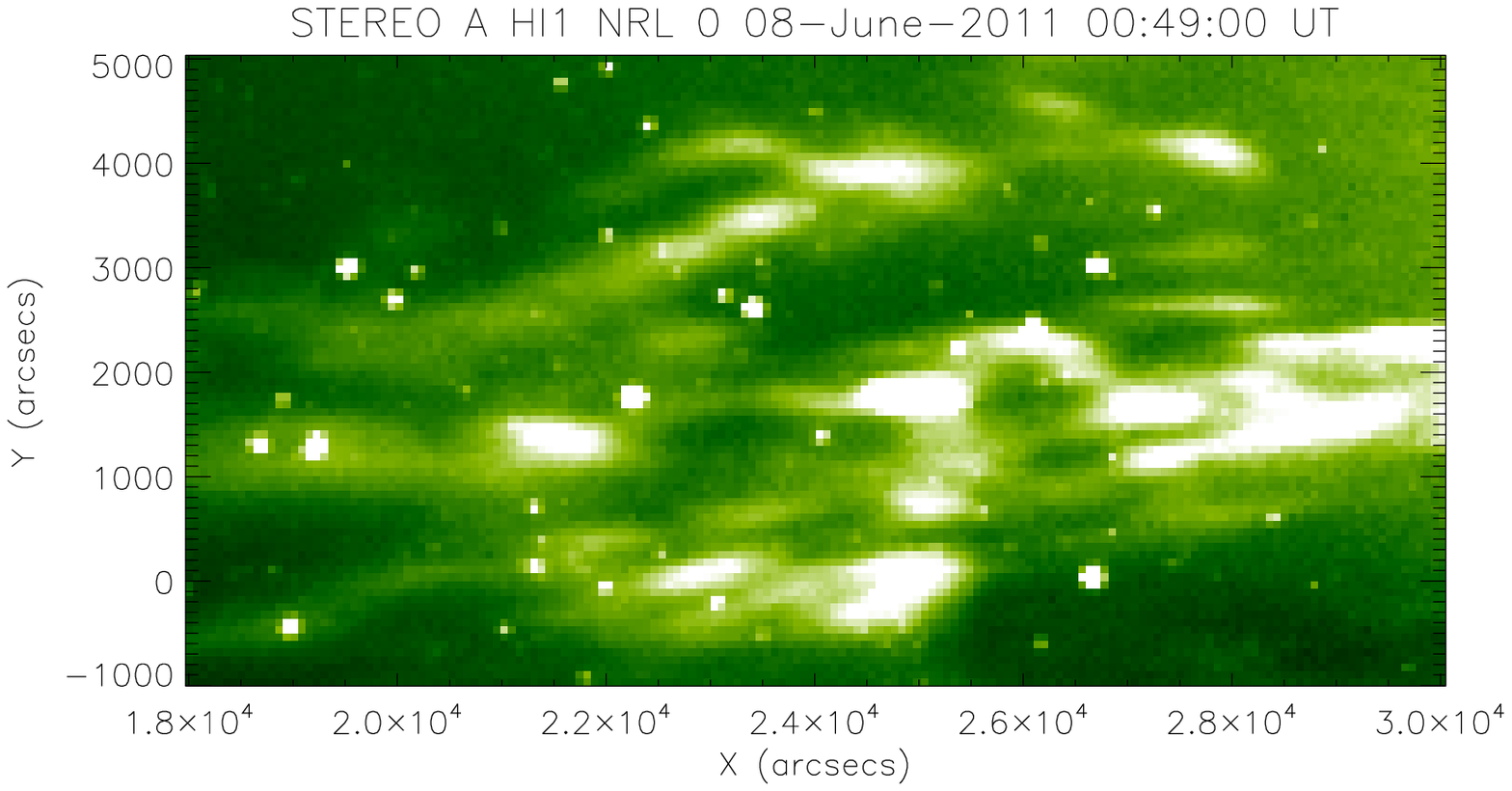}
\hspace{-1.2cm}
\includegraphics[scale=0.8,angle=0,width=9.5cm,height=10.0cm,keepaspectratio,trim=5mm 5mm 5mm 5mm,clip]{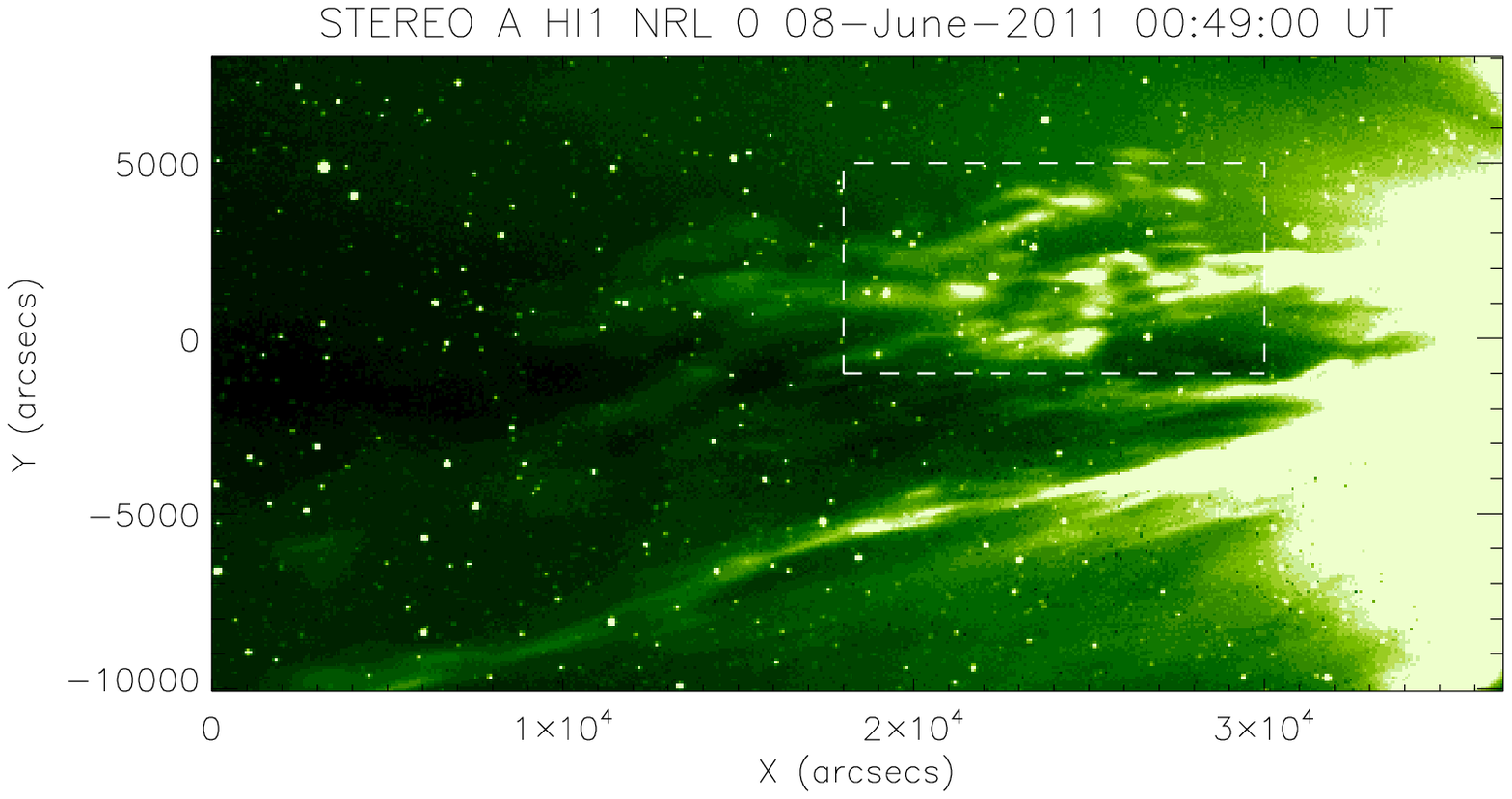}}
\end{figure*}

Initially in the intermediate corona the magnetic Rayleigh-Taylor instability grew in the linear phase, because the perturbation amplitude may be lower than the characteristic wavelength. In the outer corona where the density falls rapidly, the perturbation amplitude may be of the order of characteristics wavelength,  therefore, the linear phase may slow down and non-linear phase of an MRT instability may grew more rapidly (Sharp et al. 1984; Youngs et al. 1984). In the observed upper northward plasma segment, the magnetic Rayleigh-Taylor unstable bright finger structures are injected from an intermediate corona to the outer corona. The mushroom-like structures (M1-M5; Fig.~4) may be grown due to the breaking and twisting of some fingers of MRT unstable plasma segment, when it enters in weaker magnetic field regions in the outer corona (Youngs et al. 1984, Jun et al. 1995). 

We also observe that this plasma segment is further expanded into low inter-planetary space as observed by STEREO-A/HI-1 (Fig.~5). It is clearly visible that the instability reaches beyond 20 solar radii with respect to Sun's center. In the lower inter-planetary region, the magnetic field is weaker than outer corona, therefore the plasma fingers/mushroom-shaped unstable regions do not retain there shape and displaced to the equilibrium conditions. These structures are  fragmented, into the bright spikes (cf., Fig.~5). This fragmentation is governed by the turbulent mixing of two fluid (Sharp et al. 1984) and disappear into the inter-planetary region between 15-20 solar radii (Fig.~5; left panel).

 When we analyze the lower southward plasma segment (Fig.~2) in the outer corona, we do not find any signature of the evolution of mushroom-like structures there. This describes the fact that different phases of MRT instability highly depend upon the local plasma, magnetic field conditions, formation of the density interfaces, and finally on the physical nature of the perturbation.

\section{Discussion and Conclusions}
An unbound prominence is erupted at $7^{th}$ June 2011, and some of its part has been reached upto low inter-planetary region. The time and spatial scale were large for this eruption, and the MRT instability evolved in the plasma. In the present paper, we extend the work of Innes et al. (2012) who observed the MRT instability and related plasma structures (e.g., fingers, spikes, horns) in the same eruptive prominence in the inner corona.

We observed the part of this eruptive prominence in form of the plasma segments in STEREO-A/COR-1, COR-2, HI-1 FOV between 2.8-20 solar radii. The denser finger-like plasma structures became evident in COR-1 FOV on the upper northward plasma segment that consist of the MRT instability. The density diagnostics of these unstable fingers in the upper northward plasma segment show that heavy plasma (finger with higher density) is lying above the lighter dark plasma regions (less denser) and supported against the gravity. Linear stability theory has been used to diagnose the magnetic field required to suppress the instability which lie between 21-40 mG in this plasma segment in the intermediate corona. The estimation of the magnetic fields is of the order of (milli-Gauss) which is typical of the intermediate solar corona.
 Another MRT unstable structure, horn (H) has been clearly observed during the downfall of the plasma material into the intermediate corona in the lower southward plasma segment of the prominence. It should be noted that some falling plasma blobs are also evident in the lower part of the northward plasma segment, however, there is no Horn-like structure evident there.

We analyze the magnetic Rayleigh-Taylor instability in the upper northward plasma segment into the outer corona also (15 solar radii) as observed by STEREO-A/COR-2. The growth of the short wavelength MRT mode is fastest in comparatively strong field regions, thus initially the instability is dominated by bubbles and fingers on smaller spatial scales in the intermediate corona upto 4 solar radii. At large spatial scales, the highly twisted mushroom-like structures (M1-M5) are clearly visible in the outer corona (Fig.~4). These mushroom-like structures may be evolved due to breaking and twisting of fingers of MRT unstable plasma segment, when it enters in weaker magnetic field regions in the outer corona (Youngs et al. 1984, Jun et al. 1995).\\
As the MRT unstable regions are evolved through non-linear processes, the magnetic field decreases in the lower inter-planetary region and further converts the mushroom-like structure into localized plasma spikes (Isobe et al. 2005,2006). Similar situation is evident in the observed plasma segment containing mushroom-like MRT structures which decay further into smaller spike-like structures into the low inter-planetary space  (Fig.~5). When these structures enter into  a weak magnetic field of the inter-planetary space, they are fragmented into small-scale spikes. This may be due to the turbulent mixing of the plasma and energy exchange at smaller spatial scales (Youngs et al. 1984, Jun et al. 1995)\\
The present work, to the best of our knowledge, provides first evidence of the evolution of MRT instability upto an outer corona and low inter-planetary space. We notice that the instability features are changing shape i.e., (fingers $\rightarrow$ mushroom-like $\rightarrow$ small spikes) while magnetic field is continuously getting weaker with the height in the solar atmosphere. Such instabilities affect the transport of mass and energy in the solar atmosphere where they evolve. 

\section{Acknowledgements}
We acknowledge the constructive comments of referee that improved the manuscript. AKS and SKM acknowledge the DST-SERB (YSS/2015/000621) project. AKS acknowledges the RESPOND-ISRO (DOS/PAOGsIA2015-16/130/602) project. Authors Acknowledge the use of Dolei et al. (2014) method developed by S. Dolei to measure the density in the intermediate corona. They also acknowledge the STEREO-A \& B COR-1, COR-2 and HI-1 observational data. We thank Prof. Leon Ofman, GSFC-NASA, for discussion. PK acknowledges the support of project fund from National Science Center, Poland (Grant No: 2014/15/B/ST9/00106).

\end{document}